\documentclass[10pt, conference, a4paper]{IEEEtran}

\usepackage[english]{babel}

\IEEEoverridecommandlockouts

\usepackage{import}
\usepackage[]{units}
\usepackage{nameref}
\usepackage{url}
\usepackage{subfigure}
\usepackage{array}

\usepackage{xcolor}

\newcommand{\mbps}[1]{\unit[#1]{Mbps}}
\newcommand{\fkey}[1]{(#1)}

\ifCLASSINFOpdf
  \usepackage[pdftex]{graphicx}
\else
\fi

\usepackage[cmex10]{amsmath}

\usepackage[absolute,showboxes]{textpos}

\TPMargin{5pt}

\usepackage[inline]{enumitem}

\usepackage[np]{numprint}
\npstyleenglish

\usepackage{amssymb}

\begin{document}
	
\bstctlcite{IEEEexample:BSTcontrol}



%
\title{HASBRAIN: \\Neural Networks for HTTP Adaptive Streaming}

\title{Towards Machine Learning-Based Optimal HAS}

\author{
\IEEEauthorblockN{Christian Sieber, Korbinian Hagn, Wolfgang Kellerer}
    \IEEEauthorblockA{Chair of Communication Networks,
    \\Technical University of Munich, Germany
    \\
    \{c.sieber, k.hagn, wolfgang.kellerer\}@tum.de}
\and
    \IEEEauthorblockN{Christian Moldovan, Tobias Ho{\ss}feld}
    \IEEEauthorblockA{Chair for Modeling of Adaptive Systems,
    \\University of Duisburg-Essen, Germany
    \\
    \{christian.moldovan, tobias.hossfeld\}@uni-due.de}
}

\maketitle

\begin{abstract}

Mobile video consumption is increasing and sophisticated video quality adaptation strategies are required to deal with mobile throughput fluctuations.
These adaptation strategies have to keep the switching frequency low, the average quality high and prevent stalling occurrences to ensure customer satisfaction.
This paper proposes a novel methodology for the design of machine learning-based adaptation logics named HASBRAIN.
Furthermore, the performance of a trained neural network against two algorithms from the literature is evaluated.
We first use a modified existing optimization formulation to calculate optimal adaptation paths with a minimum number of quality switches for a wide range of videos and for challenging mobile throughput patterns.
Afterwards we use the resulting optimal adaptation paths to train and compare different machine learning models.
The evaluation shows that an artificial neural network-based model can reach a high average quality with a low number of switches in the mobile scenario.
The proposed methodology is general enough to be extended for further designs of machine learning-based algorithms and the provided model can be deployed in on-demand streaming scenarios or be further refined using reward-based mechanisms such as reinforcement learning.
All tools, models and datasets created during the work are provided as open-source software.
\end{abstract}


%
\IEEEpeerreviewmaketitle


\section{Introduction} \label{sec:introduction}

Internet video is becoming the dominant form of multimedia entertainment. 
According to the Cisco Visual Networking Index from 2017, IP video traffic will grow from 73 percent in 2016 to 82 percent of all consumer Internet traffic by 2021. 
Furthermore, live video is expected to grow 15-fold from 2016 to 2021. 
Not only does the share of Internet video traffic increase, also do the expectations of consumers grow in regard with the video quality. 
High quality and seamless delivery is expected even in difficult scenarios while stalling events during video playback have become unacceptable.
Many video service providers use adaptive streaming technologies in order to allow the client to de- or increase the bit-rate of the video during playback.
A lower video bit-rate leads to much fewer stalling events which are the largest cause for user dissatisfaction \cite{hossfeld2012initial}.
However, a lower video bit-rate leads to lower video quality and thus worse Quality of Experience (QoE). 
Therefore it is necessary to not stay on a low quality level for too long and increase the video bit-rate timely, if network conditions allow it. 
However, user studies have found that frequent bit-rate switching may have a negative impact on the QoE \cite{zink2005layer,yitong2013study}. 
Thus, we require balanced adaptation algorithms that do not switch too often while avoiding stalling and playing a video in high quality.

Many existing approaches for adaptation rely on simple heuristics which do not lead to an optimal solution. 
In these heuristics stalling can not always be avoided and the available network resources are not fully used. 

In this paper, we introduce a novel methodology for the training of machine learning-based adaptation algorithms named \textit{HASBRAIN}.
With the help of a quadratic program we generate millions of pairs of current state and the corresponding adaptation decision for the training of classification models.
The quadratic program determines the optimal adaptation path for a given video and throughput behavior.
In detail, the problem of optimal adaptation is tackled by a combined formulation of the quadratic programs discussed in our previous works \cite{hossfeld2015identifying, moldovan2017keep}. 
Similar to the original formulation of this program in \cite{miller2013optimal}, the new formulation is a two-step approach which avoids stalling events while maximizing the video quality in the first step. 
In the second step, the number of quality switches are minimized while keeping the video quality at almost maximum levels. 
Using real network traces and real videos, we are then able to calculate the optimal solution on how to adapt the video quality during playout. 

Following the sample generation, we feed the samples into multiple classification models and evaluate their performance by simulating playback sessions with the resulting trained models.
Our research questions can be formulated as follows:
\begin{itemize}
\item How close can trained models get to the optimal adaptation in HTTP adaptive video streaming?
\item How well does the trained models perform compared to other adaptation algorithms in different scenarios?
\end{itemize}
The key performance indicators that we consider include the switching frequency, the average quality, the average buffer level, the stalling frequency and the stalling time ratio. 
From the different evaluated models we present the results of a neural network in detail, as the neural network performed the best among the evaluated models.
In a nutshell, the contribution of the paper is the HASBRAIN methodology for the training of HAS adaptation algorithms with the optimal adaptation path as reference and evaluation of the resulting models.
The methodology is general enough to be extended with further learning approaches and can be used to study the relationship between current system state and adaptation decision in HAS.

The used data sets, trained models, developed tools including the simulation framework are available as open source in the supplemental material to the paper \cite{supplementalmaterial}.
The remainder of this paper is structured as follows. 
In the subsequent section we discuss background on HTTP adaptive video streaming, machine learning and Quality of Experience. 
Section \ref{sec:methodology} presents the general methodology of our work. 
In the following section we present the optimization problem for adaptive video streaming and its formulation as a quadratic program. 
Section \ref{sec:hasbrain} presents the HASBRAIN methodology in detail, including the input features and training results of different machine learning techniques.
The subsequent section \ref{sec:evaluation} presents the results of the performance evaluation.
Section \ref{sec:conclusion} summarizes the key results of the paper and discusses their implications for future work.

\section{Background \& Related Work} \label{sec:background} 

In this section HTTP Adaptive Streaming (HAS) is discussed and the state of the art regarding HAS adaptation algorithms is introduced. 
Two state of the art threshold-based adaptation algorithms for HAS, KLUDCP and TRDA, are presented.
In the performance evaluation, the two algorithms are compared to the trained models and to the optimal adaptation path.
Afterwards, the used machine learning techniques and recent approaches of HAS using reinforcement learning are briefly discussed, followed by the definition of Quality of Experience.

\subsection{HTTP Adaptive Streaming}

HTTP Adaptive Streaming (HAS) is the state of the art method for multimedia streaming. 
The most well-known HAS technique for dynamic adaption is MPEG-DASH and was first specified in 2011 by \cite{sodagar2011mpeg}. 
A HAS client consists of a buffer, a player and an adaptation logic to select the quality level. 
When starting a new streaming session, the client requests a Media Presentation Description (MPD) file from the HTTP server. 
This MPD holds information about the multimedia content that is to be streamed. 
A video is in context of HAS is split up into individual sub parts, called segments. 
The MPD gives insight into the length of a video, its individual segment sizes and segment lengths, as well as playout times of each segment and available quality representations or bit-rates.
The adaptation logic represents the decision logic that decides on the choice of the quality, i.e. bit-rate, of the first and following segments to download, given the input of the client state and other metrics such as throughput measurements.

The adaptation logic can vastly differ in the way how it selects the next quality level of a segment. 
There are threshold-based adaptation logics which select a quality level solely on the current measured throughput or buffer level.
More complex adaptation techniques consider more conditions and implement QoE optimization strategies.

\subsection{Quality of Experience}

The QoE is defined as the "degree of delight or annoyance of the user of an application or service." \cite{brunnstrom2013qualinet}. 
In video streaming it is impacted by the application layer Quality of Service (QoS) which depends on the network QoS \cite{mok2011measuring}. 
Service providers compete to provide the best QoE for users.

In user studies, it was found that stalling has a higher impact on the QoE than the initial delay of a video \cite{hossfeld2012initial,hossfeld2013internet} or a reduction in the frame rate \cite{huynh2008temporal}. 
Similarly, users are more sensitive to stalling than to an increase of a quantization parameter in the video encoder, especially for lower values of the quantization parameter \cite{singh2012quality}. 
Furthermore, stalling at irregular intervals is worse than periodic stalling \cite{huynh2008temporal}. 
A detailed analysis of the impact of network and application parameters on stalling can be found in \cite{maki2015layered}. 
In order to avoid stalling, the video bit-rate can be reduced during playtime so that video content can be downloaded into the buffer at a higher rate. However, this leads to a degradation in the video quality which has a significant impact on the QoE, as discussed in \cite{zink2005layer, yitong2013study}. 
Furthermore, it is argued that a high frequency of bit-rate adaptation events can be annoying for the user and that the bit-rate amplitude of switches should be kept low \cite{zink2005layer,yitong2013study}.
With a high buffer, switches can be delayed for longer time periods and stalling can be avoided easily. 
However, a high buffer means that more data is wasted when the user abandons the video. 
This is a trade-off that has to be considered when designing adaptive streaming algorithms.

\subsection{Adaptation Algorithms}

Next we look at the state of the art of threshold- and learning-based adaptation algorithms.
We conclude by discussing the relationship of our work with the state of the art.

\subsubsection{Threshold-Based Adaptation Algorithms}

One of the widest used adaptation algorithms are the threshold-based adaptation algorithms. 
This set of algorithms determines the quality level based on input metrics exceeding a predefined threshold. 
For example, a metric can be the application level throughput (goodput). 
If this throughput exceeds the bit-rate of the next higher quality level a threshold-based algorithm may decide to switch to this next higher quality representation. 
Another commonly used metric is the current buffer level and is often used in addition to the throughput metric. 
When more metrics are used algorithms allow for a more complex adaptation behavior. 
Two examples for threshold-based adaptation algorithms are KLUDCP \cite{muller2012evaluation} and TRDA \cite{miller2012adaptation}. 
These algorithms are used in this work to characterize the performance of our adaptation approach.
KLUDCP is known for being an aggressive adaptation algorithm with high average quality and a high quality switching frequency, while TRDA is known for being conservative with a medium average quality and a low switching frequency \cite{sieber13implementation}.

\subsubsection{Machine Learning-Based Algorithms}

QoE-aware HAS approaches based on reinforcement learning have been studied recently by researchers. 
These approaches focus on the use of Q-learning, a reinforcement learning technique that can be used to find an optimal action in a Markov decision process by learning to find an optimal action in a state of the system following the optimal policy.
This approach was first described and discussed by \cite{watkins1992q}.

In \cite{claeys2013design} a Q-learning client is trained for a HAS scenario considering a QoE model to reward the client for the playout of high average playout quality while punishing extensive switching. 
This approach is further refined in \cite{claeys2014design} by including the state of the buffer fullness to diminish the effect of high volatile throughput pattern and its performance is evaluated in \cite{claeys2014design2}. 
Further improvements of this approach in terms of multi-user fairness are presented in \cite{van2015learning} and evaluated in  \cite{martin2016evaluation}. 

A recent approach of the use of reinforcement learning for HAS is shown in \cite{mao2017neural}. 
To create an adaptation algorithm an ANN is trained with a state of the client-side network and video player measurements, i.e. the measured bandwidth, the buffer fullness and the playback bit-rate, and classifies the next segments bit-rate to download. 
Hereby the ANN is rewarded when full-filling certain QoE metrics such as the playout quality as a function of the playout smoothness, the perceived quality improvement and the maximization of the highest playout quality. 
The ANN is penalized when an action leads to re-buffering. 
This approach outperforms existing algorithms by 12\% to 25\% according to the authors of the paper.

We circumscribe ourself from the state of the art by centering our methodology and evaluation around the optimal adaptation path.
The optimal adaptation path shows how to adapt the quality optimally under a given trade-off between average quality and switching frequency.
By learning from the optimal adaptation path, the switching frequency can be kept low and the average quality high, even in scenarios with highly varying throughput.
Machine learning models trained in such a way can be used as they are in streaming clients or further re-fined using reinforcement learning.

\section{Methodology} \label{sec:methodology}  

In the following we first give an overview over our work.
Afterwards we discuss the goodput dataset, the training and validation content and the discrete event simulation used for the performance evaluation in detail.

\begin{figure}[t]
\centering
\includegraphics[width=0.99\columnwidth]{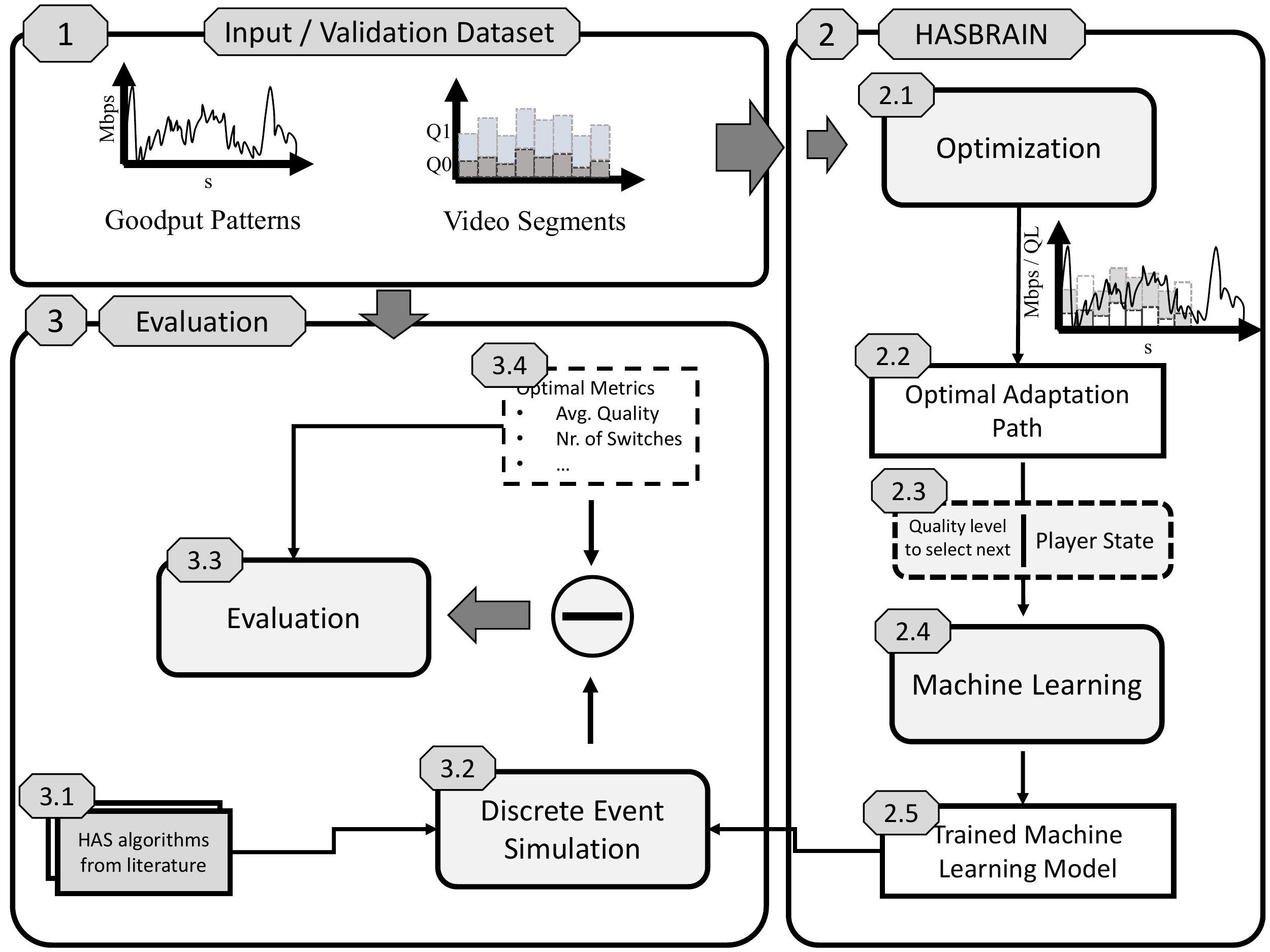}
\caption{
Overview over the design and evaluation of HASBRAIN. 
}
\label{fig:overview}
\end{figure}

Figure \ref{fig:overview} illustrates the work divided into the input and validation datasets \fkey{1}, the model training approach denoted as HASBRAIN \fkey{2} and the performance evaluation \fkey{3}.
The input datasets consist of a challenging mobile goodput traffic pattern and video content.
The goodput pattern describes the application-level downstream throughput recorded while driving on a highway in Austria.
The pattern exhibits frequent fluctuations and drops in available goodput.
The content dataset consists of video segments taken from a popular online video platform divided into content for the training and validation.

In \fkey{2}, the dataset is fed into an optimization formulation which calculates the optimal adaptation path based on a trade-off between switching frequency and average quality.
From the resulting adaptation path we extract the decision points and the player state at the time instants of the decisions as training samples \fkey{2.3}.
Hence, a training sample consists of the player state, e.g. current buffer level, observed throughput, known segment sizes, segment size variations, etc., and the corresponding optimal quality level decision with respect to the calculated optimal adaptation path.

In \fkey{2.4}, we take the extracted samples to train different machine learning models such as SVM and an artificial neural network.
For the evaluation we subsequently select the model which performs best on the training samples \fkey{2.5}.

For the evaluation we select the validation content and goodput patterns and simulate the client adaptation behavior with the trained machine learning model and two algorithms taken from the literature \fkey{3.1} \& \fkey{3.2}.
Additionally we calculate the optimal adaptation path for each combination of content and goodput pattern.
Afterwards we compare the observed behavior of the simulated algorithms with the optimal adaptation path in terms of average quality, switching frequency, average buffer level and stalling events \fkey{3.3} \& \fkey{3.4}.

\subsection{Goodput Pattern} \label{subsec:goodput}


Figure \ref{fig:goodput_video_segments} depicts the goodput pattern used in the training and evaluation.
The goodput pattern describes the application-level downstream throughput recorded while driving on a highway in Klagenfurt, Austria in December 2012 \cite{sieber13implementation}.
In this work, the recorded trace is scaled to a mean of \mbps{0.67} with a coefficient of variation (CV) of 0.38.
An autocorrelation of 0.80 for a lag of \unit[1]{s} is observed.
The duration of the pattern is \unit[720]{s} with a sample frequency of \unit[1]{Hz}.
For the training and validation we chose goodput starting points $t$ from a set of 101 starting points $t \in \{0, 7, 14, ..., 700\}$.
Is the end of the goodput pattern reached, the goodput pattern restarts from $t = 0$.
By default, one training/validation sample has a memory of $c_M$=30 seconds. 
This results in 24 unique throughput memory vectors and in total 720 different vectors where some vectors have overlapping sequences

The pattern represents a challenging mobile scenario with frequent drops and unstable throughput.
Adaptation algorithms have to either be conservative and aim for a high buffer level, adapt frequently to the new throughput, try to predict the future throughput based on the throughput observed so far, or a combination of the three options.

\subsection{Content Characteristics} \label{subsec:content}

\begin{figure}[t]
\centering
\includegraphics[width=0.92\columnwidth]{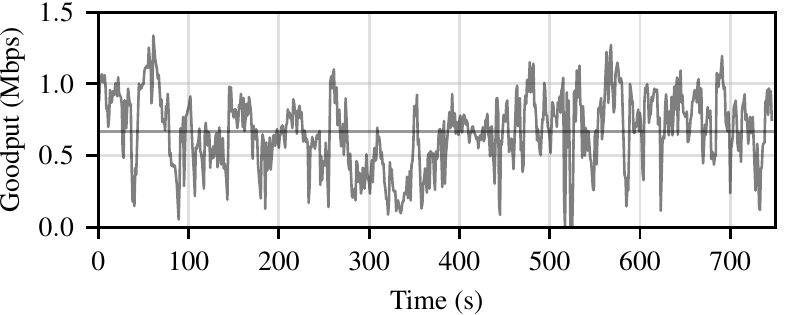}\\
\includegraphics[width=0.92\columnwidth]{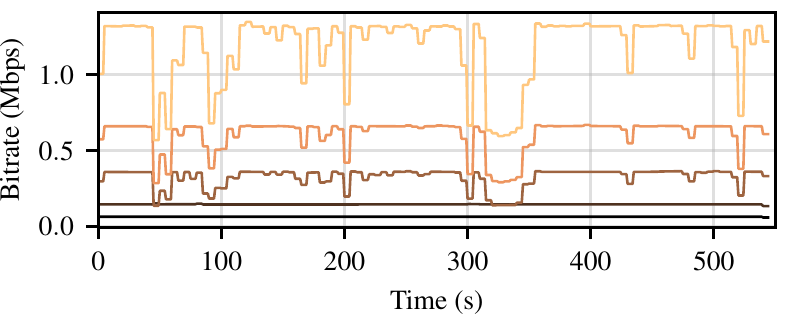}\\
\caption{
Challenging mobile goodput pattern (top) used in the training and evaluation and average bit-rate over time for one of the selected YouTube videos (bottom) for the quality levels $144p$, $240p$, $360p$, $480p$ and $720p$.
}
\label{fig:goodput_video_segments}
\vspace{-1em}
\end{figure}


The content for the training and validation is randomly selected from YouTube.
For training set we select 41 videos with durations of 1 to 10 minutes (average \unit[5.3]{minutes}).
We select videos where five quality levels ($144p, 240p, 360p, 480p, 720p$) are available.
The average bit-rate of the quality levels are \mbps{0.1}, \mbps{0.23}, \mbps{0.36}, \mbps{0.68} and \mbps{1.33}.
\cite{sieber16sacrificing} gives further details on the selection process, characteristics of the videos and tools used.
We segment the videos to a constant segment duration of \unit[1]{s}.

For the validation set we select 20 additional videos with durations ranging from 4 to 8 minutes (average \unit[5.79]{minutes}).
The average bit-rates of the videos range from \mbps{0.06} for $144p$ to \mbps{0.99} for $720p$.
Figure \ref{fig:goodput_video_segments} depicts the average bit-rate for one of the videos in the training set. 
The behavior of bit-rate over time suggests that the video is encoded with a maximum bit-rate for each quality level.
The drops in bit-rate indicate scenes of low motion or low level of detail.
In the example video, the drop can be as much as from \mbps{1} to zero at around \unit[510]{s}.

The encoding behavior illustrates why it can be beneficial to a HAS adaptation algorithm to consider bit-rate variations for its decisions.
In the DASH standard, the manifest file allows to include information for all segments of the video.
Hence, the algorithm knows how the video bit-rate will behave in the future and can make decisions accordingly.
For example, the algorithm may decide during a period of low bit-rate to increase the quality level and decrease it right before the bit-rate increases again.
Average bit-rate, often used by heuristics as a decision criteria, is inadequate to capture the characteristics of the videos.

\subsection{Discrete Event Simulation} \label{subsec:simulation}

We use a Discrete Event Simulation (DES) to evaluate the performance of the adaptation algorithms with respect to a specific goodput pattern and content.
The simulation takes three inputs: The algorithm to use and its parameters, the segment sizes of the quality levels of a video and a goodput pattern.
The simulation does not apply any random variations on the input.
This allows us to deterministically evaluate the performance of the algorithms on the exact same inputs as the calculation of the optimal adaptation path uses.
To validate this, we implemented an adaptation algorithm which can access the future goodput behavior and which calculates the optimal adaptation path.
Simulation runs with the decisions from the optimal adaptation path result in the exact output as calculated.
The simulation framework is provided as part of the supplemental material to the paper \cite{supplementalmaterial}.

\section{Optimal Adaptation}
\label{sec:optimal_adaptation}

\begin{table}[]
\centering
\caption{General Notation}
\label{tab:genvars}
\begin{tabular}{l|p{6cm}}
 Variable & Definition\\
 \hline
 $r\: [5]$  & Available representations\\ 
 $n\: [200]$ & Number of segments \\ 
 $\tau\:[1]$ & Duration of a segment in [s] \\ 
 $S_{ij}$ & Size of a segment i from representation j \\ 
 $D_i$ & Playback deadline for segment i \\ 
 $T_0\: [5]$ & Start-up (or initial) delay in [s]\\ 
 $V(t)$ & Total amount of data V(t) received by a client during the time [0,t]\\
 $T(v)$ & Time $T(v)$ required by a client to download volume $v$. $T(v)$ is the inverse function of $V(t)$, i.e. $T(V(t))\: =\: t$\\
 $x_{ij}\:\in \{0,1\} $&Target variable indicating if client downloads segment i from representation j ($x_{i j}\: =\: 1$) or not ($x_{i j}\: =\: 0$)\\
 \hline 
\end{tabular}
\end{table}

In order to create a training set for the training of the models, we first formulate an optimization problem. 
The optimal solution to the optimization problem is then used to extract training samples.
We chose the optimization problem as first suggested in \cite{miller2013optimal} and modified in \cite{hossfeld2015identifying}.
The problem in their approach is that they first optimize the mean quality, and then reduce the number of switches as far as possible. 
This does not leave much room for optimization of the latter since the number of switches can not be reduced significantly.
Previous work \cite{liotou2016enriching, metzger2016tcp} suggests combining the two-step approach into a single step with weights for the mean quality and for the number of switches. 
The single step approach was then used in \cite{moldovan2017keep} to investigate the trade-off between the mean quality and the number of switches in adaptive video streaming. 
By combining the two-step approach into a single step with weights for the mean quality and for the number of switches, it was found that by reducing quality by a very small margin, the number of switches can be reduced significantly most of the time. 


In this paper, we improve the two-step approach by an optimality gap parameter $\epsilon$ that aims at balancing both QoE values at reasonable levels. 
This addition to the two-step approach allows to decrease the number of switches drastically while the mean quality is only reduced by a small margin $\epsilon$. 


In the first step, we calculate the highest mean quality $W_{opt}$ that can be achieved while avoiding stalling. Equation \ref{eq:11} and \ref{eq:12} dictate that each segment is downloaded in exactly one representation. Equation \ref{eq:13} regulates that each segment $k$ is downloaded before its deadline $D_k$. Furthermore, it says that the sum of the data of all downloaded segments may not be larger than the data that can be downloaded until that point in time.
~
\begin{align}
\label{obj1}\mathrm{maximize} \;\;\;& W_{opt}=\frac{1}{n} \sum^n_{i=1} \sum^r_{j=1} jx_{ij} \\
\label{eq:11}\mathrm{subject\,to} \;\;\;& \sum_{j=1}^rx_{ij}=1 \;\; \forall i \in \{1,\dots,n\}\\
\label{eq:12}& x_{ij}\in \{0,1\} \;\; \forall i \in \{1,\dots,n\},\; j \in \{1,\dots,r\}\\
\label{eq:13}&\sum_{i=1}^{k}\sum_{j=1}^rS_{ij}x_{ij}\leq V(D_k) \;\; \forall k \in \{1,\dots,n\}
\end{align}

In the second step, we find an optimal adaptation so the number of switches is minimized while the mean quality is at least $W_{opt}-\epsilon$ while avoiding stalling. Equation \ref{eq:21}-\ref{eq:23} are identical to Equation \ref{eq:11}-\ref{eq:13}. Equation \ref{eq:24} limits the mean quality of all downloaded segments to the the value that was determined in the first step minus an $\epsilon$ that leaves room for the minimization of switches.

\begin{align}
\label{obj2}\mathrm{minimize} \;\;\;& \frac{1}{2} \sum^{n-1}_{i=1} \sum^r_{j=1} (x_{ij} - x_{i+1,j})^2 \\
\label{eq:21}\mathrm{subject\,to} \;\;\;& \sum_{j=1}^rx_{ij}=1 \;\; \forall i \in \{1,\dots,n\}\\
\label{eq:22}& x_{ij}\in \{0,1\} \;\; \forall i \in \{1,\dots,n\},\; j \in \{1,\dots,r\}\\
\label{eq:23}&\sum_{i=1}^{k}\sum_{j=1}^rS_{ij}x_{ij}\leq V(D_k) \;\; \forall k \in \{1,\dots,n\}\\
\label{eq:24}&\frac{1}{n} \sum^n_{i=1} \sum^r_{j=1} jx_{ij} \geq W_{opt} - \epsilon, \epsilon \in \{0,\dots r\}
\end{align}

\section{Model Training} \label{sec:hasbrain}

In this section we describe the machine learning models, parameters and training in detail.
The goal of the model is to learn from the optimal adaptation path which quality level to pick for the next segment.
It is a classification problem where the model has to deduce which quality level to pick next based on the past and current player state, previous decisions, goodput measurements and video characteristics.

The optimization is able to calculate the optimal adaptation path for a given goodput pattern and content as it is given the full goodput pattern beforehand.
But an adaptation logic in the player does not know how the goodput changes in the future.
The player's decision criteria are limited to measurements of the current goodput and previous measurements.
However, the model can learn to make good decisions based on the observed goodput behavior.
For example, the model may look at the goodput measurements and classify the goodput as unstable in the last seconds.
The optimal adaptation path tells the model what a good decision is for the current player state and goodput, such as: Be conservative, sudden drops in available goodput are likely.
We denote the inputs based on previous decisions or measurements as memory inputs.
The memory, video and player state features and scaling approaches are defined in detail in Section \ref{subsec:input}, together with the output labels.
Section \ref{subsec:mltech} introduces the used machine learning techniques and why for the final evaluation the artificial neural network was chosen.
Section \ref{subsec:rebuf} discusses the implemented rebuffering strategy in cases where stalling occurred.

\subsection{Input Features, Scaling, Ouput Labels} \label{subsec:input}

Next we discuss the features of the input samples and the feature scaling.
Table \ref{tab:input_features} summarizes the features and scaling variables.
The features are divided into the three categories \textit{memory features}, \textit{video features} and \textit{player state features}.

\textit{Memory features} store the last states of selected features.
The memory of the selected quality levels stores the last $c_M$ quality levels chosen by the optimal adaptation path.
The memory of the observed goodput stores the observed goodput of the last $c_{OT}$ segments.
With the memory features the machine learning model can relate the current situation to previous decisions and states.
The \textit{video feature} future representations represents the segment sizes of the next $c_M$ segments relative to the current playback position of all available quality levels as provided by the DASH MPD file.
This gives the machine learning model the possibility to react to future drastic changes in the video bit-rate such as spikes due to fast-motion or drops due to low motion scenes.
In the category \textit{player state features}, we define the current buffer level as input feature.

In total there are 242 input features for the default value of 30 for $c_{M}$ and $c_{OT}$.
For the throughput there are 30 input features from the observed throughput over the last $c_{OT}$ segments and 1 from the average of this observed throughput. 
For the memory inputs, there are 30 features from the memory of the selected quality levels and again 30 for the memory of the selected segment sizes. 
150 further input features are from the segment sizes over all 5 available quality levels for the next $c_{M}$ future segments. 
The last feature is the current bufferlevel.

\begin{table}[]
\centering
\caption{HASBRAIN Input Features}
\label{tab:input_features}
\begin{tabular}{p{1cm}p{4cm}}
\multicolumn{1}{l|}{Definition} & Description \\ \hline
\multicolumn{2}{c}{Memory Features}                                          \\ \hline
\multicolumn{1}{l|}{$ \frac{1}{\nu} \sum \limits_{k=1}^{c_{OT}} \frac{Tp_{d_{i-c_{OT}+k}}}{c_{OT}} $}           & Observed average throughput of the last $c_{OT} $ segments.          \\
\multicolumn{1}{l|}{$\frac{1}{\nu}\cdot\{ {Tp_{{d_{i-c_{OT}}}},...,Tp_{{d_{i}}}}\} $}           & Observed throughput for the download of the last $c_{OT}$ segments.             \\
\multicolumn{1}{p{2.2cm}|}{$\frac{x_{(i-k)j}\cdot j}{r} $ $,(\forall x|x_{ij}=1)$ $,\forall k\in\{c_{M},...,1\}  $}           & Memory of selected quality levels            \\ 
\multicolumn{1}{p{2.2cm}|}{$\frac{S_{(i-k)j}}{\nu} $  $,(\forall x|x_{ij}=1),\forall k\in\{c_{M},...,1\} $}           & Memory of selected segments            \\ \hline

\multicolumn{2}{c}{Video Features}                                           \\ \hline
\multicolumn{1}{p{3.2cm}|}{$ \frac{S_{(i+k)j}}{\nu}$ $,\forall k \in\{0,...,c_{M}\}$ $\forall j \in\{1,...,r\} $}           & Segment sizes of the $c_M$ future segments of all available representations \\ \hline
\multicolumn{2}{c}{Player State Features}                                    \\ \hline
\multicolumn{1}{l|}{$ \frac{1}{Bl_{max}}\cdot Bl_{d_{i}}$}           & Buffer level at time $ t\: =\: t_{d_i}$ normalized by the maximum Buffer level            \\ \hline 
\multicolumn{2}{c}{Other Symbols}                                    \\ \hline
\multicolumn{1}{l|}{$\nu = \smash{\displaystyle\max_{\forall i,j}}(Tp_{d_i},S_{ij})$ \vspace{0.6em}}   & Bit-rate/Throughput scaling         \\ \hline 
\multicolumn{1}{l|}{$Bl_{max}$ [20]}   & Buffer level scaling \\ \hline  
\multicolumn{1}{l|}{$c_{OT}$ [30]}   & Memory of observed goodput  \\ \hline 
\multicolumn{1}{l|}{$c_{M}$ [30]}  & Segments to consider  \\ \hline 
\end{tabular}
\end{table}

%

We standardize all input features by scaling the features to a range of zero to one.
Video bit-rate and throughput are scaled using the same variable to preserve the relationship between them.
$\nu$ denotes the bit-rate and throughput scaling variable and is defined as the maximum video bit-rate or throughput value observed during our experiments ($\smash{\displaystyle\max_{\forall i,j}}(Tp_{d_i},S_{ij})$).
$Bl_{max}$ denotes the maximum buffer-level allowed by the client and is used to scale the past and current buffer level observations.

In the beginning of a video streaming session there are not enough input features available to represent the memory inputs. 
These features are set to a fixed value of 0. 
Similarly, towards the end of the streaming session there are not enough video features to represent the future video input. 
These features are set to a value of 0 as well.

The output of the machine learning model is defined as $y_{x} | \forall x \in \{0,..,r-1\}$ with $y_x \in [0, 1]$ being the probability of quality level $x$ to be chosen for the next segment to download.
The quality level with the highest probability is chosen by the client.


\subsection{Machine Learning Techniques} \label{subsec:mltech}

Three different machine learning techniques were considered for the evaluation: Support Vector Machines (SVM), k-Nearest Neighbors and Artificial Neural Networks (ANN).

In the following we describe the design and training of the ANN.
The ANN consists of an input, one hidden and one output layer.
The layers are full-connected and the sigmoid activation function is chosen for all input and hidden layer neurons. 
This activation function maps an input of variable range to an output in the range of $[0,1]$.
The input layer consists of 242 neurons as defined in the previous Section \ref{subsec:input}.
For the hidden layer, 110 neurons where chosen.
The output layer consists of $r$ neurons, one for each quality level to chose ($y_{x} | \forall x \in \{0,..,r-1\}$).
The resulting labels are chosen by the softmax function which maps the input to a categorical probability distribution with $r$ different outcomes.
TensorFlow \cite{abadi2016tensorflow} was used for the implementation, training and deployment of the ANN in this work.

The SVM is configured to use a radial basis function kernel. 
By applying this kernel trick it allows this linear classifier method to map the non linear feature input into a a transformed feature space and fit the maximum-margin hyperplane on this transformed features.

For the k-Nearest Neighbors technique, the k value of nearest features that are used to determine the class is set to 5. 
The weights of the distances is set to be uniform, so that every distance is weighted the same, i.e. with a value of 1.

These techniques are trained with a sample set of around 900.000 samples. 
This training set is split up into 800.000 training samples and 100.000 validation samples to determine the learning success. 
The 100.000 validation samples are \textit{not} from the validation video set, but also from the training video set.
Here we are interested in the performance of the training, not how the trained model performs as a HAS adaptation algorithm in the simulation.

The learning success is assessed with the classification accuracy. 
k-Nearest Neighbors achieved on this set an accuracy of 97.69\% whereas the SVM achieved 98.74\%. 
The NN outperformed both other techniques with an accuracy of 99.1\% and with roughly 1 hour of training has the shortest training time compared to over 10 hours for the other two approaches.
Due to this achieved high validation accuracy of the ANN, only this technique was further considered for the evaluation.

One might also use Recurrent Neural Networks (RNN) and especially in the form of a Long Short Term Memory (LSTM) networks. 
Those approaches were deliberately not further explored in this work as the learning performance on the samples is already over \unit[99]{\%}.
We argue that the left over decisions are of unpredictable nature due to the fact that the optimal adaptation has knowledge of the future goodput.
However, future work should nevertheless investigate LSTM and RNN networks on the samples for completeness.



\subsection{Rebuffering Strategy} \label{subsec:rebuf}

We use a rebuffering strategy called D-policy from \cite{moldovan2016bridging} with a D value of 10s. 
This means if a stalling event occurs the video playout is halted until at least 10s of video segments to playout are in the buffer or the last segment to download has arrived.  
This strategy was shown to allow an adaptation algorithm to better cope with challenging scenarios like the one used in this paper.


\section{Evaluation} \label{sec:evaluation}  

We evaluate the performance of the trained ANN, TRDA and KLUDCP on the validation video set in terms of the switching frequency, average playback quality, the average buffer level, the frequency of stalling events and the stalling ratio.
The trained ANN is denoted has HASBRAIN in the figures.

\subsection{Evaluation Methodology} \label{subsec:evaluation_meth}

Next we give details on the evaluation methodology and sequence of events in the evaluation.
First, a video is picked from the validation content.
Second, a starting point for the goodput pattern is chosen from the aforementioned set of 101 starting points and all goodput samples up to the starting point are appended to the end of the pattern.

Third, the video and the resulting goodput pattern are used as input for the simulation.
The simulation uses the trained model, the KLUDCP algorithm and the TRDA algorithm to simulate one playback session with each of the algorithms with the provided video and goodput pattern.
Should the end of the pattern be reached before the end of the playback session, the pattern is restarted from the beginning.
In parallel, the goodput pattern and video is used as input for the optimization and the optimal adaptation path is calculated.

The playout of the video begins after the initial delay of \unit[5]{s} as defined in Table \ref{tab:genvars}. 
If there is not at least one fully downloaded segment in the playout buffer after the delay is expired then the playout is further delayed until this requirement is met.
The memory features are initialized with zeros as described in Section \ref{subsec:input}.
An exploratory study showed that the performance of the ANN during the first and last minute of the streaming session is not different than during the session.

After the streaming session, the optimal adaptation path and the output of the simulation is summarized using the same metrics, e.g. switching frequency, average quality, etc.
The sets of metrics of the algorithms are then compared to the optimal adaptation path metrics and new differential metrics are created, e.g. difference in switching frequency, difference in average quality, etc.

The process is repeated for all 20 videos in the validation set and for every of the 101 starting points, which results in roughly 2000 simulation runs per algorithm.

\subsection{Switching Frequency}

\begin{figure}[t]
\centering
\subfigure[Switching Frequency]{\includegraphics[width=0.47\columnwidth]{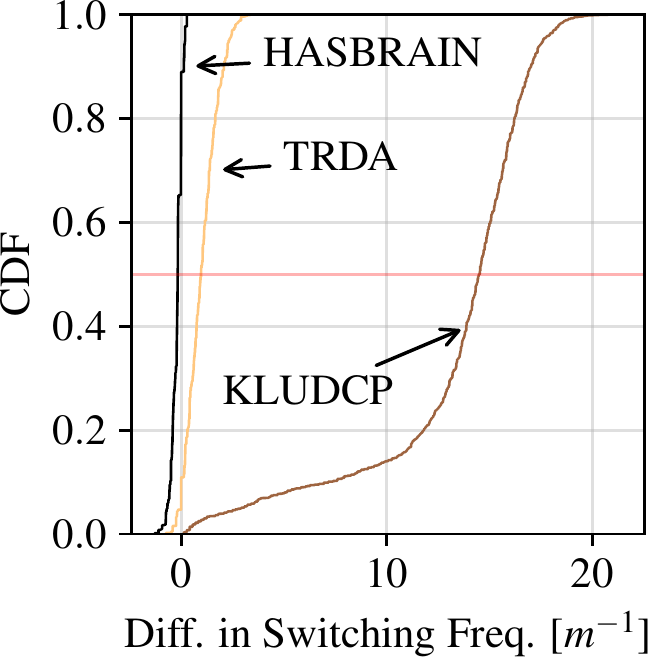}\label{fig:02_switching_frequency}}
\subfigure[Average Quality]{\includegraphics[width=0.47\columnwidth]{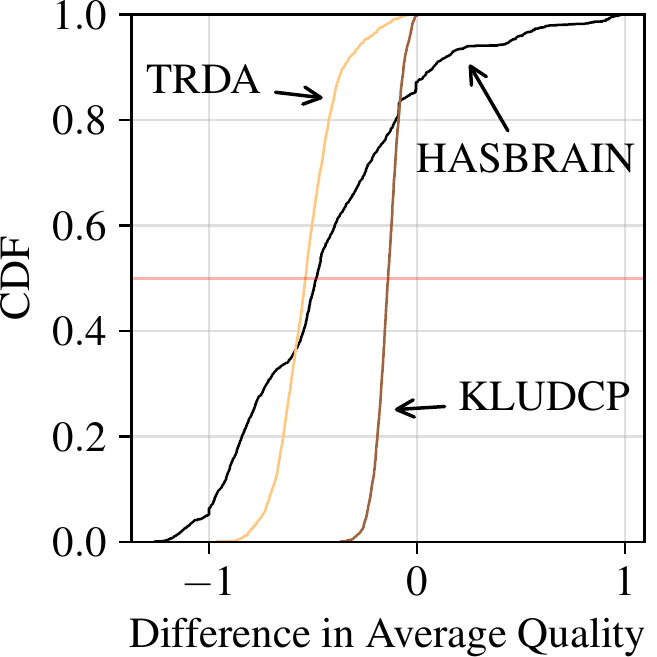}\label{fig:03_avg_quality}}\\
\caption{
Difference in switching frequency of the three investigated algorithms compared to the optimal adaptation path as CDF. 
HASBRAIN is trained with the optimal adaptation path and exhibits a comparable switching behavior. 
}
\label{fig:stalling}
\vspace{-1em}
\end{figure}

Figure \ref{fig:02_switching_frequency} shows the difference in switching frequency in terms of switches per minute ($m^{-1}$) compared to the optimal adaptation path over all simulation runs as a cumulative distribution function (CDF).
The horizontal line indicates the median.
Up to \unit[89]{\%} of the HASBRAIN runs exhibit a lower or equal switching frequency than the optimal adaptation.
The left over \unit[11]{\%} of the runs do not surpass the optimal adaptation path by \unit[0.3]{$m^{-1}$}.
For TRDA, roughly \unit[10]{\%} of the runs exhibit an equal or less switching frequency than the optimal adaptation.
The other \unit[89]{\%} surpass the optimal adaptation by up to 0.35 per minute.
For KLUDCP, none of the simulation runs exhibit a lower switching frequency than the optimal adaptation.
Up to \unit[20]{\%} of the runs stay below an increase of 12 switches per minute.
A maximum of up to 21.4 more switches per minute can be observed for KLUDCP.

From the figure we conclude that the HASBRAIN is very conservative in terms of switching and can keep the switching frequency as low as the optimal adaptation path.
TRDA, which is also described in the literature as conservative \cite{sieber13implementation}, switches in \unit[89]{\%} of the runs more often than the optimal adaptation, but half the runs stay below an increase of 1 switch more per minute.
KLUDCP, described as aggressive in previous studies \cite{sieber13implementation}, exhibits a switching frequency increase by up to 21 switches per minute.

\subsection{Average Quality}


Figure \ref{fig:03_avg_quality} shows the difference in average quality between the three evaluated algorithms for the validation runs.
Note that an increase in average quality compared to the optimal adaptation path is possible (positive values in the figure).
This is due to the fact that the $\epsilon$ in the optimization formulation leaves a (tiny) room of improvement \cite{moldovan2017keep}.
Furthermore, the algorithms in the simulation are allowed to stall the playback and therefor have more time to download video segments, which is impossible for the optimization, which is not allowed to stall.

The figure shows that in general KLUDCP exhibits an average playback quality close to the optimal adaptation.
The maximum decrease in average quality is roughly 0.4 quality levels and \unit[50]{\%} of the runs are only 0.1 quality levels worse than the optimal adaptation.
TRDA's maximum decrease is 0.9 quality levels and \unit[50]{\%} of the runs are not worse than 0.55 quality levels.
From this it follows that the average playback quality with TRDA is on average roughly half a quality level lower than with KLUDCP.
HASBRAIN shows mixed results over all simulation runs.
For \unit[13]{\%} of the simulation runs, the playback quality is increased compared to the optimal adaptation, up to one quality level.
On the other side, \unit[50]{\%} of the simulation runs exhibit a by 0.5 quality levels lower average quality than the optimal adaptation.
For some runs, the average playback quality is 1.3 quality levels lower than the optimal adaptation.
The median difference is located between TRDA and KLUDCP with about half a quality level.

The results for the difference in average quality shows expected behavior for the relative performance of TRDA and KLUDCP.
For HASBRAIN, it shows that there are some runs with disproportional low and high average quality.

\subsection{Average Buffer Level}


The average buffer level is the time-dependent average of the playback buffer level in terms of seconds during the whole video playback session of a simulation run.
On the one hand, a high buffer level allows the video player to compensate for sudden decreases in available throughput.
On the other hand, a low buffer level decreases the time between a quality level decision for a video segment and the time the segment is shown to the user.
In the simulation framework, the buffer space is assumed to be unlimited.

\begin{figure}[t]
\centering
\includegraphics[width=0.97\columnwidth]{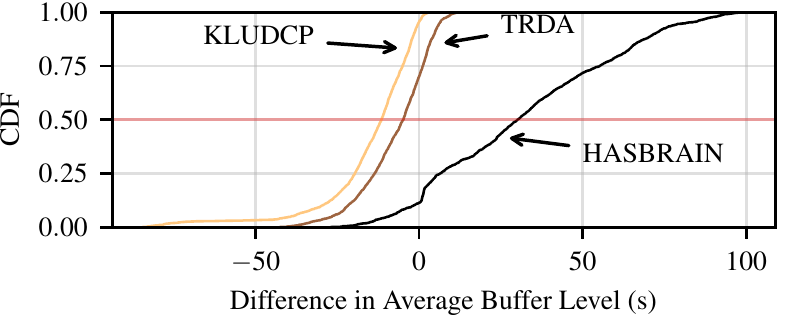}
\caption{
Difference in the time-dependent average of the buffer level during the simulation runs compared to the optimal adaptation path. 
}
\label{fig:04_avg_buff_level}
\end{figure}

Figure \ref{fig:04_avg_buff_level} illustrates the difference in average buffer level for all runs for the three evaluated algorithms compared to the optimal adaptation path.
The simulation results show that \unit[96]{\%} of KLUDCP and \unit[70]{\%} of TRDA runs exhibit a lower average buffer level than the optimal adaptation path.
The runs which exhibit an increase in average buffer level can increase it up to \unit[12]{s} and \unit[5]{s} seconds, respectively.
HASBRAIN increases the average buffer level for \unit[89]{\%} of the runs, while only \unit[11]{\%} of the runs show a lower average buffer level.
The maximum increase of \unit[100]{s} and a maximum decrease of \unit[27]{s} is observed over all runs.

The results show that HASBRAIN keeps the buffer level high for most runs, even compared to the conservative optimal adaptation path.
KLUDCP and TRDA both exhibit a lower average buffer level.

\subsection{Stalling Frequency}

Next we discuss the results regarding the stalling frequency of the three algorithms.
As the optimal adaptation path is by definition without stalling events, it is not included in the comparison.

Stalling frequency describes the number of stalling events per minute.
A stalling event occurs when the playback buffer is empty and the playback has to be paused until a certain amount of playback time is available again as defined by the re-buffering strategy.
Although the quantitative impact on the QoE is debated, a stalling event has a negative affect on the experience of the user and should be avoided \cite{hossfeld2012initial}.

\begin{figure}[t]
\centering
\subfigure[Stalling Frequency]{\includegraphics[width=0.47\columnwidth]{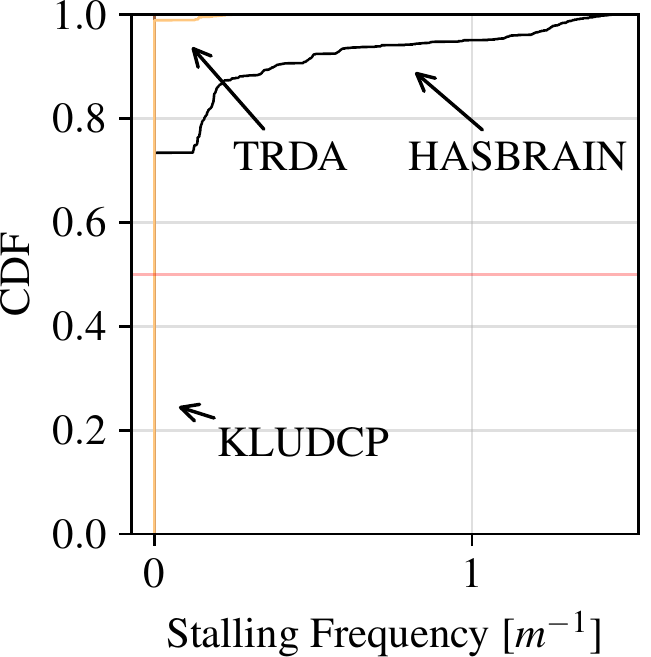}\label{fig:05_stalling_freq}}
\subfigure[Stalling Time Ratio]{\includegraphics[width=0.47\columnwidth]{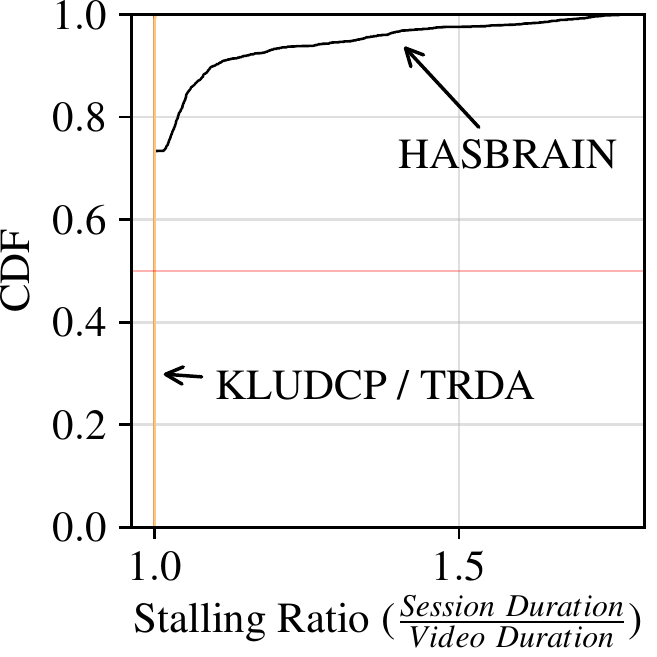}\label{fig:06_stalling_ratio}}\\
\caption{
Comparison of the stalling events the three investigated algorithms as CDF over all validation runs. 
}
\label{fig:stalling}
\end{figure}

Figure \ref{fig:05_stalling_freq} shows the stalling frequency over all simulation runs for the three algorithms as CDF.
The results show that with KLUDCP there is no stalling event in all of the simulation runs.
With TRDA, only \unit[1]{\%} of the runs exhibit any stalling with a maximum of 3 stallings per minute.
HASBRAIN exhibits no stalling for \unit[73]{\%} of the runs.
From the \unit[27]{\%} which exhibit stalling, the stalling frequency is less than once in three minutes.
For the other runs, the maximum observed stalling frequency is 1.45 stallings per minute.

From the results follows that KLUDCP and TRDA are good at preventing stalling events in the investigated scenario.
HASBRAIN can prevent stalling events in three quarter of the simulation runs.
However, there are some runs which exhibit stalling events.

\subsection{Stalling Time Ratio}

Next we take a look at the stalling time ratio.
We define the stalling time ratio as the ratio between the duration of the whole playback session and the duration of the video.
For example, a stalling time ratio of 1.5 for a 30 second video states that there was a cumulative duration of 15 seconds where the playback paused.
For algorithms with a low initial buffering time and no stalling events during the playback, the stalling ratio is close to 1.

Figure \ref{fig:06_stalling_ratio} shows the stalling time ratio as CDF over all simulation runs.
As KLUDCP and TRDA exhibit no stalling events or only very few stalling events, respectively, the stalling ratios of all runs with the two algorithms are close to 1.
For HASBRAIN, \unit[73]{\%} of the runs show a stalling ratio of close to 1.
For the other \unit[27]{\%}, a maximum stalling ratio of 1.76 is observed.

The stalling occurrences is due to two reasons. 
First, the optimal paths can rely on the knowledge about the future throughput to make perfect decisions even when the buffer level is low.
The trained model can only rely on the previously observed throughput and estimations how it may behave during the consecutive seconds.
Hence, some decisions are not understandable by the model and can lead to wrong decisions, especially when the buffer level is already low.
Second, the learning approach does not include explicit costs or rewards for stalling occurrences or for preventing stallings, respectively.

\section{Discussion \& Conclusion} \label{sec:conclusion}  

HTTP-based adaptive video streaming makes up an increasing portion of today's mobile Internet traffic.
But especially mobile connections are prune to frequent throughput fluctuations and thus require sophisticated video quality adaptation strategies.
Hence, to ensure customer satisfaction with the video service, quality adaptation algorithms have to balance quality switching frequency, average quality and the risk of stallings.
This paper proposes a novel methodology for the design of machine learning-based adaptation logics named HASBRAIN.
A modified version of an existing optimization formulation is used to calculate optimal adaptation paths with minimal number of switches in a challenging mobile scenario.
The optimal adaptation paths are used to train different machine learning models and based on the training performance, an artificial neural network is selected for further evaluation.
Furthermore, the trained artificial neural network is compared against two existing adaptation algorithms.

In a nutshell, the evaluation shows that the ANN keeps quality switching frequency low and the average buffer level high.
In terms of average quality, the ANN shows a higher average quality in a majority of runs compared to the conservative TRDA and a lower average quality than the aggressive KLUDCP.
However, for some runs the ANN exhibits a lower average quality than TRDA and a higher average quality than KLUDCP.
The ANN is not able to avoid stalling completely and stalls in \unit[5]{\%} of the runs more than once a minute.
However, the results of the ANN in this challenging mobile scenario are promising.
Reinforcement learning is a candidate for further fine-tuning of the trained ANN model to reduce the stalling occurrences to zero while at the same time keeping the quality and switching frequency close to the performance of the optimal adaptation.


The provided methodology and open source code can be used by the community for the development and evaluation of machine learning-based adaptation strategies.
The trained neural network can be used as adaptation logic or utilized as a seed for reward-based learning techniques such as deep reinforcement learning.

\section*{Acknowledgements}
This work was partly funded  by  the  German  Research Foundation (DFG) under the grant number KE1863/6-1. 
It was also partly funded by Deutsche Forschungsgemeinschaft (DFG) under grants HO 4770/1-2 (DFG \"Okonet: Design and Performance Evaluation of New Mechanisms for the Future Internet -- New Paradigms and Economic Aspects).
The authors alone are responsible for the content of the paper.

\bibliographystyle{IEEEtran}

\bibliography{bibliography}

%
%
%


\end{document}